\documentclass[twocolumn,preprintnumbers,amsmath,amssymbm,prl]{revtex4}
\usepackage{epsfig}
\usepackage{graphicx}

\begin{document}
\title{Return of the quantum cosmic censor}
\author{Shahar Hod}
\address{The Ruppin Academic Center, Emeq Hefer 40250, Israel}
\address{ }
\address{The Hadassah Institute, Jerusalem 91010, Israel}
\date{\today}

\begin{abstract}
\ \ \ The influential theorems of Hawking and Penrose demonstrate
that spacetime singularities are ubiquitous features of general
relativity, Einstein's theory of gravity. The utility of classical
general relativity in describing gravitational phenomena is
maintained by the cosmic censorship principle. This conjecture,
whose validity is still one of the most important open questions in
general relativity, asserts that the undesirable spacetime
singularities are always hidden inside of black holes. In this
Letter we reanalyze extreme situations which have been considered as
counterexamples to the cosmic censorship hypothesis. In particular,
we consider the absorption of {\it fermion} particles by a spinning
black hole. Ignoring quantum effects may lead one to conclude that
an incident fermion wave may over spin the black hole, thereby
exposing its inner singularity to distant observers. However, we
show that when quantum effects are properly taken into account, the
integrity of the black-hole event horizon is irrefutable. This
observation suggests that the cosmic censorship principle is
intrinsically a quantum phenomena.
\end{abstract}
\bigskip
\maketitle


Spacetime singularities that arise in gravitational collapse are
always hidden inside of black holes, invisible to distant observers.
This is the essence of the weak cosmic censorship conjecture (WCCC),
put forward by Penrose forty years ago \cite{HawPen,Pen,Haw1,Brady}.
The validity of this hypothesis is essential for preserving the
predictability of Einstein's theory of gravity. The conjecture is
based on the common wisdom that singularities are not pervasive
\cite{Brady} and it has become one of the cornerstones of general
relativity. Moreover, it is being envisaged as a basic principle of
nature. However, despite the flurry of research over the years, the
validity of this conjecture is still an open question (see e.g.
\cite{Wald1,Sin,Clar,Vis,Price,Wald2,His,KayWal,BekRos,Hub,QuiWal,Hod1,HodPir,Hod2,ForRom1,ForRom2,MatSil,Hodever,RicSaa}
and references therein).

The destruction of a black-hole event horizon is ruled out by this
principle because it would expose the inner singularities to distant
observers. Moreover, the horizon area of a black hole, $A$, is
associated with an entropy $S_{BH}=A/4$ \cite{Beken1} (we use
natural units in which $G=c=\hbar=1$). Therefore, without any
obvious physical mechanism to compensate for the loss of the
black-hole enormous entropy, the destruction of the black-hole event
horizon would violate the generalized second law (GSL) of
thermodynamics \cite{Beken1}. For these two reasons, any process
which seems, at first sight, to remove the black-hole horizon is
expected to be unphysical. For the advocates of the cosmic
censorship principle the task remains to find out how such candidate
processes eventually fail to remove the horizon.

According to the uniqueness theorems \cite{un1,un2,un3,un4,un5}, all
stationary solutions of the Einstein-Maxwell equations are uniquely
described by the Kerr-Newman metric which is characterized by three
conserved parameters: the gravitational mass $M$, the angular
momentum $J$, and the electric charge $Q$. A black-hole solution
must satisfy the relation
\begin{equation}\label{Eq1}
M^2-Q^2-a^2 \geq 0\  ,
\end{equation}
where $a\equiv J/M$ is the specific angular momentum of the black
hole. Extreme black holes are the ones which saturate the relation
(\ref{Eq1}). As is well known, the Kerr-Newman metric with
$M^2-Q^2-a^2<0$ does not contain an event horizon, and it is
therefore associated with a naked singularity rather than a black
hole.

One may try to ``over spin" a black hole by injecting into it
particles with small energy and large angular momentum. In this work
we inquire into the physical mechanism which protects the black-hole
horizon from being eliminated by the absorption of waves which may
``supersaturate" the extremality condition, Eq. (\ref{Eq1}). In
order to analyze such processes one should study the propagation and
scattering of various fields in the black-hole spacetime.

The dynamics of a wave field $\Psi$ in the rotating Kerr-Newman
spacetime is governed by the Teukolsky equation \cite{Teu,Dud}. One
may decompose the field as
\begin{equation}\label{Eq2}
\Psi_{lm}(t,r,\theta,\phi)=e^{im\phi} {_sS_{lm}}(\theta;a\omega)
_s\psi_{lm}(r;a\omega)e^{-i\omega t}\  ,
\end{equation}
where $(t,r,\theta,\phi)$ are the Boyer-Lindquist coordinates,
$\omega$ is the (conserved) frequency of the mode, $l$ is the
spheroidal harmonic index, and $m$ is the azimuthal harmonic index
with $-l\leq m\leq l$. The parameter $s$ is called the spin weight
of the field, and is given by $s=\pm 2$ for gravitational
perturbations, $s=\pm 1$ for electromagnetic perturbations, $s=\pm
{1\over 2}$ for massless neutrino perturbations, and $s=0$ for
scalar perturbations. (We shall henceforth omit the indices $l,m$,
and $s$ for brevity.) With the decomposition (\ref{Eq2}), $\psi$ and
$S$ obey radial and angular equations, both of confluent Heun type
\cite{Heun,Flam}, coupled by a separation constant $A(a\omega)$
\cite{NoteAlm}.

For the scattering problem one should impose physical boundary
conditions of purely ingoing waves at the black-hole horizon and a
mixture of both ingoing and outgoing waves at infinity (these
correspond to incident and scattered waves, respectively).
Namely,
\begin{equation}\label{Eq3}
\psi \sim
\begin{cases}
e^{-i\omega y}+{\mathcal{R}}(\omega)e^{i \omega y} & \text{ as }
r\rightarrow\infty\ \ (y\rightarrow \infty)\ ; \\
{\mathcal{T}}(\omega)e^{-i (\omega-m\Omega)y} & \text{ as }
r\rightarrow r_+\ \ (y\rightarrow -\infty)\ ,
\end{cases}
\end{equation}
where the ``tortoise" radial coordinate $y$ is defined by
$dy=[(r^2+a^2)/\Delta]dr$, with $\Delta\equiv r^2-2Mr+Q^2+a^2$. [The
zeroes of $\Delta$, $r_{\pm}=M\pm (M^2-Q^2-a^2)^{1/2}$, are the
black hole (event and inner) horizons.] Here $\Omega=a/(r^2_++a^2)$
is the angular velocity of the black hole. The coefficients ${\cal
T}(\omega)$ and ${\cal R}(\omega)$ are the transmission and
reflection amplitudes for a wave incident from infinity.

The transmission and reflection amplitudes satisfy the usual
probability conservation equation $|{\cal T}(\omega)|^2+|{\cal
R}(\omega)|^2=1$. The calculation of these scattering amplitudes in
the low frequency limit, $M\omega\ll 1$, is a common practice in the
physics of black holes, see e.g. \cite{Chan,Page} and references
therein. For {\it boson} fields ($s=0,\pm1,\pm2$) one finds
\cite{Page,Hodever}
\begin{eqnarray}\label{Eq4}
|{\cal T}(\omega)|^2\sim {{\omega-m\Omega}\over{\pi
T_{BH}}}(AT_{BH}\omega)^{2l+1}\  ,
\end{eqnarray}
for the transmission probability, where $T_{BH}=(r_+-r_-)/A$ is the
Bekenstein-Hawking temperature of the black hole, and
$A=4\pi(r^2_++a^2)$ is its surface area.

The transmission probability, Eq. (\ref{Eq4}), implies that those
modes for which the frequency $\omega$ and the azimuthal quantum
number $m$ are related by $\omega<m\Omega$ have negative
transmission probabilities. These modes are actually amplified
rather than absorbed. This is the well-known black-hole
superradiance phenomena \cite{PreTeu,Beko}. One therefore finds that
only modes for which
\begin{equation}\label{Eq5}
\omega>m\Omega\  ,
\end{equation}
can be absorbed by the black hole. Thus, it is impossible to
increase the black-hole spin without increasing its mass
simultaneously. This fact guarantees that the black-hole condition,
Eq. (\ref{Eq1}), would still be respected after the absorption of
the mode \cite{Hodever,Noteever}. One therefore concludes that the
incident mode cannot remove the black-hole horizon. Cosmic
censorship is therefore respected!

We have just learned that, thanks to the superradiance phenomena the
black-hole is protected from being over spinned by an incident {\it
bosonic} mode. It should be emphasized, however, that the same
cannot be said if the incident mode is of a {\it fermion} type. It
is a well-known fact that there is no superradiance effect for
fermion fields \cite{Unr1,Unr2,Guv,WaDa,Goul,RicSaa}. That is, the
superradiant term $\omega-m\Omega$ is absent from the expression of
the transmission probability of fermion fields. For fermion fields
one simply finds \cite{Page,RicSaa}
\begin{eqnarray}\label{Eq6}
|{\cal T}(\omega)|^2\sim (AT_{BH}\omega)^{2l+1}\ .
\end{eqnarray}
It may therefore seem, at first sight, that fermion particles of
low-energy and high angular momentum are not hindered from entering
the black hole. This has led Richartz and Saa \cite{RicSaa} to
conclude that an incident fermion wave may over spin the black hole,
thereby exposing its inner singularity to distant observers.

Everything in our past experience in physics tells us that a black
hole should defeat any attempt of destroying its event horizon. It
seems every time we think we have finally found a sophisticated way
of violating the cosmic censorship conjecture, nature still has the
final word. The cosmic censorship principle asserts that nature
should always provide a black hole with some physical mechanism
which would protect its integrity, thereby preventing one from
exposing the black-hole inner singularity.

Where should we look for the physical mechanism which may protect
the cosmic censorship principle in the current case ? The absence of
superradiance for fermion fields is a direct consequence the Pauli
exclusion principle (for fermion fields there can only be one
quantum per state). Being a quantum principle, it suggests that
there may be some {\it quantum} phenomena which protects the
integrity of the black-hole horizon in the present gedanken
experiment.

Indeed, it has been shown by Unruh \cite{Unr1,Unr2} that the
intrinsic parity non-conservation of neutrinos (i.e., massless
neutrinos have only one helicity) would lead to vacuum polarization
effects about a spinning black hole. What is the physical cause for
this polarization effect? In the rotating Kerr spacetime there is a
gravitational spin-orbit coupling between orbiting particles and the
black-hole spin. Near the black hole (more precisely, inside the
ergosphere), the spin-orbit coupling becomes strong enough to create
negative energy orbits (as seen from infinity) \cite{Car}. The
phenomena of vacuum polarization in the Kerr spacetime is a direct
consequence of the possibility of decay into such negative-energy
orbits \cite{Unr1,Unr2}.

The vacuum polarization effect is the wave analog of the Penrose
process \cite{Penp}, in which rotational energy can be extracted
from a rotating black hole. The process utilizes the existence of
retrograde particle orbits in the ergosphere of rotating black
holes, for which the energy, as it would have been measured by an
observer at infinity, is negative \cite{Pir}. Such orbits can not
come out to infinity. However, the negative-energy particles can
induce changes in the energies of other particles, which do come out
to infinity.

The spontaneous polarization of the vacuum around rotating black
holes involves the creation of two modes. One of these modes is
co-rotating with the black hole and is characterized by a positive
energy as measured at infinity, while the other one is counter
rotating, having negative energy. The positive-energy particle can
escape to infinity, while the negative-energy particle must fall
into the black hole. An observer at infinity may detect the emitted
positive-energy particle. He also measures a decrease in the
rotational energy of the black hole, caused by the infall of the
retrograde, negative-energy particle into it. He therefore concludes
that by the spontaneous polarization process rotational energy was
extracted from the black hole \cite{Penp,Pir}.

It is worth noting that, the fact that the energy-momentum tensor of
neutrinos can have negative values inside the ergosphere
\cite{Unr1,Unr2,Guv,WaDa,Goul} is in immediate contradiction with
the positive energy condition. This energy criterion is commonly
assumed to be valid for classical matter distributions. Thus, the
spontaneous creation of particles in the Kerr spacetime is obviously
a {\it quantum} phenomena.

It should be emphasized that the quantum polarization effect of
neutrinos in the Kerr spacetime exists even for extremal,
zero-temperature black holes. In fact, taking cognizance of
Hawking's expression for the expected number of particles in each
fermion mode of the black-hole radiation \cite{Hawth}
\begin{equation}\label{Eq7}
<N_{lms}\omega>=|{\cal
T}_{lms}(\omega)|^2\{\exp[(\omega-m\Omega)/T_{BH}]+1\}^{-1}\  ,
\end{equation}
and substituting $T_{BH}=0$ for the extremal limit, one finds
\begin{equation}\label{Eq8}
<N_{lms}\omega>=|{\cal T}_{lms}(\omega)|^2\ \Theta(m\Omega-\omega)\
,
\end{equation}
where $\Theta(x)$ is the Heaviside step function.

The result Eq. (\ref{Eq8}) implies that even cold ($T_{BH}=0$)
spinning black holes emit fermion particles with frequencies below
$m\Omega$. As indicated by Unruh \cite{Unr1,Unr2}, this vacuum
polarization effect serves to constantly decrease both the mass and
the angular momentum of the black hole. It should be emphasized that
only fermion modes for which $\omega<m\Omega$ are being
spontaneously emitted by the extremal spinning black hole. This
implies that the black hole loses angular momentum more rapidly than
it loses its (squared) mass. One therefore concludes that the black
hole is constantly pushing itself away from the extremal limit [that
is, the dimensionless ratio $(r_+-r_-)/r_+$ increases with time due
to the spontaneous emission of neutrinos].

The important point to realize is that the Pauli exclusion principle
would prevent one from beaming low-energy neutrinos (characterized
by $\omega<m\Omega$) on the rotating black hole any more rapidly
than they are spontaneously emitted \cite{Notehigh}. Thus, the
exclusion principle implies that one can at best suppress the
constant increase of the ratio $(r_+-r_-)/r_+$. Cosmic censorship is
therefore respected!

A closely related problem is that of a rotating black hole immersed
in a thermal radiation bath \cite{Bekim}. Taking into account the
quantum character of the black hole (namely, the spontaneous
polarization of the vacuum around the black hole) one finds that the
probability $p(0|1)$ that one fermion mode is incident upon the
black hole with no reflection is given by \cite{Bekim}
\begin{equation}\label{Eq9}
p(0|1)=|{\cal T}(\omega)|^2(1+e^{-x})^{-1}\  ,
\end{equation}
where
\begin{equation}\label{Eq10}
x\equiv(\omega-m\Omega)/T_{BH}\  ,
\end{equation}
and $|{\cal T}(\omega)|^2$ is given by Eq. (\ref{Eq6}).

The absorption probability (\ref{Eq9}) implies that the claim of
Ref. \cite{RicSaa} that the probability to over spin a black hole is
given by $|{\cal T}(\omega)|^2$ [as defined in Eq. (\ref{Eq6})] is
actually erroneous. The correct probability, given by Eq.
(\ref{Eq9}), is actually suppressed by a factor of
$(1+e^{-x})^{-1}$. It is clear that in order to over spin the black
hole one must consider an incident mode characterized by
$\omega<m\Omega$ (that is, a mode with $x<0$). For extremal black
holes (characterized by $T_{BH}=0$) this implies $x\to -\infty$, and
one therefore finds from Eq. (\ref{Eq9}) that $p(0|1)=0$. The
incident mode is therefore reflected with probability $1$. In
particular, it fails to over spin the extremal black hole.

Suppose we have a near extremal black hole with mass $M$ and angular
momentum $J=M^2-1$ (this is the ``nearest extreme" black hole
considered in \cite{RicSaa}). Of course, our analysis is meaningful
only in the semiclassical regime, where $M>>1$. In this limit, the
temperature of the near-extremal black hole is given by
$T_{BH}\simeq \sqrt{2}/4\pi M^2$. Consider an incident mode with
small frequency and $l=m=3/2$. In this case $x=-3\pi M/{\sqrt 2}$.
Taking cognizance of Eq. (\ref{Eq9}) one finds that the probability
of such a `dangerous' mode to be absorbed by the black hole is
\begin{equation}\label{Eq11}
p(0|1)=|{\cal T}(\omega)|^2e^{-3\pi M/{\sqrt 2}}\  .
\end{equation}
It is obvious that the absorption probability is extremely small in
the semiclassical regime $M>>1$ considered here. For example, for a
black hole of mass $M=1$gr, the suppression factor $(1+e^{-x})^{-1}$
is $\sim 10^{-133009}$ \cite{Notesmall}. The number $10^{133009}$ is
much larger than the total number of particles in the whole
universe. It is therefore clear that the absorption probability of a
dangerous mode is practically zero.

Moreover, such dangerous absorptions (with extremely low
probability) are not cumulative-- most of the incident modes will
merely be scattered off the black hole. In being scattered they will
always radiate into the black hole some gravitational waves
\cite{Beko} which will push the black hole away from the extremal
limit.

In summary, extreme situations which have been considered as
counterexamples to the weak cosmic censorship conjecture were
reexamined. In particular, we have reanalyzed gedanken experiments
in which {\it fermion} waves are beamed from far away towards a
near-extremal rotating black hole. The unique feature which
characterizes fermion fields is the absence of the superradiance
phenomena. It therefore seems, at first sight, that fermion modes of
low energy and large angular momentum can be absorbed by the black
hole. One may thus give in to the temptation of concluding that the
black hole can acquire enough angular momentum to over spin,
$M^2-Q^2-a^2<0$. Previous analyzes \cite{RicSaa} indeed claimed that
this process may provide a counterexample to the WCCC. However, we
have demonstrated that a more complete analysis of the gedanken
experiment (in which {\it quantum} effects are properly taken into
account) reveals that it does {\it not} violate the weak cosmic
censorship conjecture.

The physical mechanism which protects the integrity of the
black-hole horizon in this gedanken experiment is the spontaneous
emission of low-energy ($\omega<m\Omega$) fermions by the rotating
black hole (quantum instability of the vacuum around a spinning
object). This vacuum polarization effect was discovered by Unruh in
the $70$'s after performing a detailed analysis of the second
quantization of fields in the rotating Kerr spacetime
\cite{Unr1,Unr2}. It is interesting to note that, historically one
could have {\it predicted} the existence of such a spontaneous
quantum emission already in the $60$'s, following the formulation of
the cosmic censorship principle. We thus conclude that the cosmic
censor must be cognizant of both general relativity and quantum
physics.

\bigskip
\noindent {\bf ACKNOWLEDGMENTS}
\bigskip

This research is supported by the Meltzer Science Foundation. I
thank Jacob D. Bekenstein, Tsvi Piran and Liran Shimshi for
stimulated discussions. I also thank Yael Oren for helpful
discussions.



\begin{thebibliography}{99}
\bibitem{NoteAlm} The functions $S(\theta;a\omega)$ are the spheroidal wave functions
\cite{Teu,Flam}. In the $a\omega\ll 1$ limit they become the
familiar spin-weighted spherical harmonics with the corresponding
angular eigenvalues $_sA_{lm}=l(l+1)-s(s+1)+O(a^2\omega^2)$.
\bibitem{Notehigh} Note that one can still send high-energy neutrinos (characterized by
$\omega>m\Omega$) into the black hole. However, such high-energy
modes would certainly push the black hole away from the extremal
limit. [That is, they will increase the ratio $(r_+-r_-)/r_+$.]
Thus, high-energy particles pose no challenge to the cosmic
censorship conjecture.
\bibitem{Notesmall} It is worth emphasizing that the pre-factor $|{\cal
T}(\omega)|^2$ is also very small, see Eq. (\ref{Eq6}).
\end{thebibliography}
\end{document}